\documentclass[10pt,twocolumn,letterpaper]{article}

\usepackage[pagenumbers]{wacv} 

\usepackage{graphicx}
\usepackage{amsmath}
\usepackage{amssymb}
\usepackage{booktabs}
\usepackage{makecell}

\usepackage[pagebackref,breaklinks,colorlinks]{hyperref}

\usepackage[capitalize]{cleveref}
\crefname{section}{Sec.}{Secs.}
\Crefname{section}{Section}{Sections}
\Crefname{table}{Table}{Tables}
\crefname{table}{Tab.}{Tabs.}

\def\wacvPaperID{2151} 
\def\confName{WACV}
\def\confYear{2025}

\begin{document}

\title{InvisMark: Invisible and Robust Watermarking for AI-generated Image Provenance}

\author{Rui Xu \and Mengya Hu \and Deren Lei \and Yaxi Li \and David Lowe \and Alex Gorevski \and Mingyu Wang \and Emily Ching \and Alex Deng \\
Microsoft Responsible AI\\
{\tt\small \{rxu,humia,derenlei,yaxi.li,david.lowe,algore,mwang,yuetc,alex.deng\}@microsoft.com }
}
\maketitle

\begin{abstract}
The proliferation of AI-generated images has intensified the need for robust content authentication methods. We present InvisMark, a novel watermarking technique designed for high-resolution AI-generated images. Our approach leverages advanced neural network architectures and training strategies to embed imperceptible yet highly robust watermarks. InvisMark achieves state-of-the-art performance in imperceptibility (PSNR$\sim$51, SSIM $\sim$ 0.998) while maintaining over 97\% bit accuracy across various image manipulations. Notably, we demonstrate the successful encoding of 256-bit watermarks, significantly expanding payload capacity while preserving image quality. This enables the embedding of UUIDs with error correction codes, achieving near-perfect decoding success rates even under challenging image distortions. We also address potential vulnerabilities against advanced attacks and propose mitigation strategies. By combining high imperceptibility, extended payload capacity, and resilience to manipulations, InvisMark provides a robust foundation for ensuring media provenance in an era of increasingly sophisticated AI-generated content. Source code of this paper is available at: \href{https://github.com/microsoft/InvisMark}{https://github.com/microsoft/InvisMark}.
\end{abstract}

\section{Introduction}
\label{sec:intro}

The rapid advancement of generative AI (GenAI) technologies has revolutionized the creation of digital images, enabling the production of hyper-realistic deepfakes with unprecedented ease. While this technological leap offers exciting possibilities for creative expression, it simultaneously poses significant challenges to information integrity and public trust. The potential for these AI-generated images to be used in manipulating elections, damaging reputations, and undermining societal foundations underscores the urgent need for robust solutions to verify the origin and authenticity of digital content~\cite{ferrara2024genai}.

In response to these challenges, the Coalition for Content Provenance and Authenticity (C2PA) has emerged as a collaborative effort aimed at combating misinformation within the digital content ecosystem~\cite{rosenthol2020content, c2pa2024}. 
The C2PA suggests adding signed provenance information directly into its metadata. However, this approach is vulnerable to metadata stripping by malicious actors or during content sharing on social media platforms~\cite{bharati2021transformation,black2021vpn, zhang2020discovering}.  In this case, the stripped provenance can potentially be recovered through soft bindings, such as fingerprinting or watermarking, from trusted repository.  Fingerprinting techniques leverage near-duplicate search in trusted databases for recovering content provenance~\cite{bharati2021transformation,black2021vpn, zhang2020discovering}. However, these matches often lack precision, necessitating human intervention for verification. Image watermarking offers an alternative solution by inserting an imperceptible identifier within the content itself~\cite{cox2007digital}. This allows for exact matching and retrieval of associated provenance from databases.

Traditional watermarking techniques have focused on embedding imperceptible patterns within images, either directly in pixel values or in transformed frequency domains. Pixel-based methods, such as least significant bit (LSB) embedding, are simple to implement but highly susceptible to removal~\cite{wolfgang1996watermark}. Frequency-domain techniques, utilizing transforms like Discrete Wavelet Transform (DWT) or Discrete Cosine Transform (DCT), offer improved robustness to certain transformations and have seen industrial adoption, such as the commerical implementation of Stable Diffusion ~\cite{al2007combined, rombach2022high}. 
Despite the improvements, frequency domain methods still suffer from vulnerability to relatively minor alterations to the image, limiting their robustness in real-world scenarios~\cite{hsieh2001hiding, holub2012designing, potdar2005survey}. 

\begin{figure*}[t]
  \centering
   \includegraphics[width=1.0\linewidth]{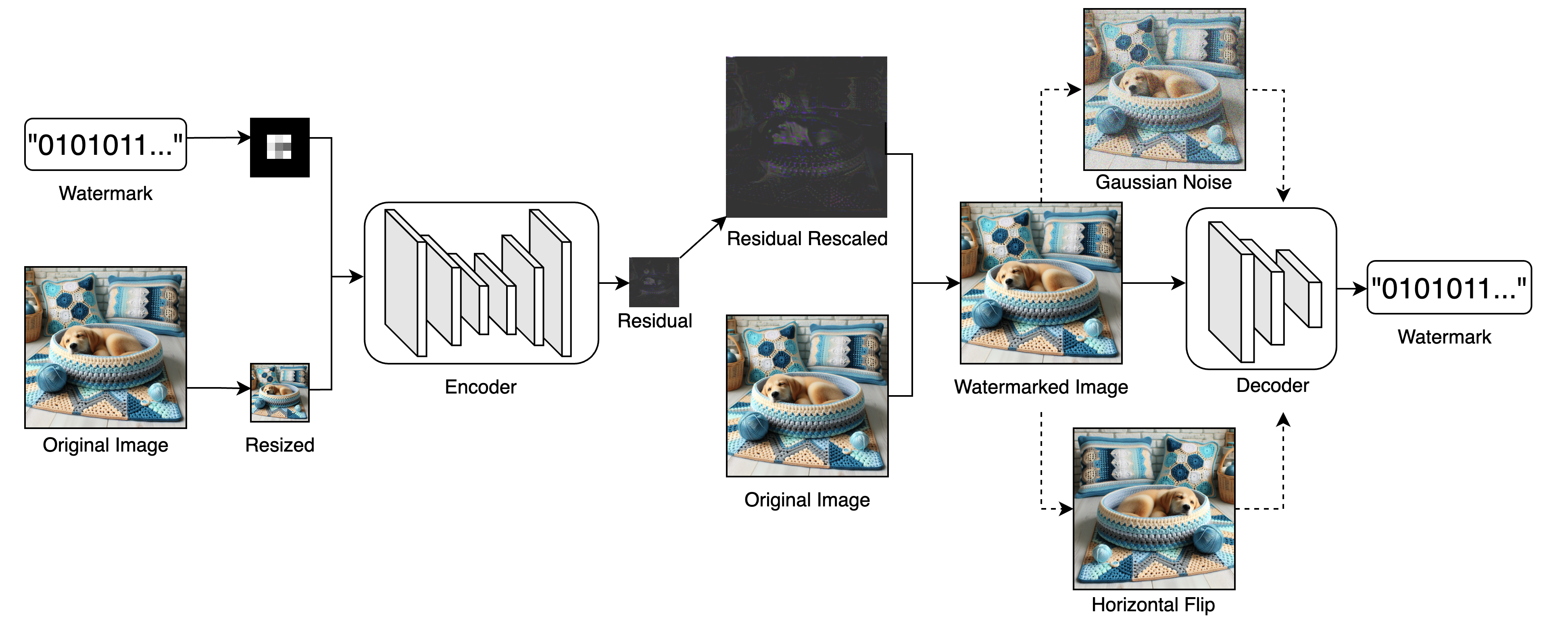}
   \caption{Overview of our method for watermark encoding and decoding. Watermark first passes through a preprocessing layer and then concatenates with the resized cover image. A MUNIT-based encoder generates watermark residuals, which are upscaled and added to the original image, producing the watermarked image. To ensure robustness, we select the top-k noises that yield the poorest watermark recovery from a pre-defined set of noises, these losses are incorporated into watermark training.}
   \label{fig:model_flowchart}
\end{figure*}

The advent of GenAI has spurred the development of innovative watermarking algorithms that integrate seamlessly into the image generation process. These approaches include watermarking training images with pre-trained encoders and decoders, followed by fine-tuning generative models on these watermarked images~\cite{yu2021artificial}. Alternatively, some methods focus on fine-tuning only the decoder of a Latent Diffusion Model (LDM) while leaving the diffusion component unchanged~\cite{lukas2023ptw}. However, these approaches are computationally intensive and often model-specific. Furthermore, their primary targeted applications have been in AI-generated image detection or user identification~\cite{saberi2023robustness, fernandez2022watermarking}, rather than content provenance tracking.


In contrast, post-generation watermarking offers greater flexibility, as it can be applied to any image regardless of its origin. Methods like HiDDen~\cite{zhu2018hidden} and RivaGAN~\cite{zhang2019robust} utilize encoder-decoder architecture to embed hidden messages within images. Subsequent works have proposed various improvements to enhance image quality and robustness~\cite{chang2021retracted, duan2019reversible, meng2018fusion, tancik2020stegastamp}. For instance, StegaStamp~\cite{tancik2020stegastamp} introduces significant image perturbations between encoder and decoder, enabling the encoded image to withstand real-world distortions. TrustMark~\cite{bui2023trustmark} proposed a scaling-based approach for watermarking images of any resolution. Beyond encoder-decoder architectures, SSL~\cite{fernandez2022watermarking}  suggests embedding watermarks within the self-supervised latent space by shifting image features into a designated area. RoSteALS ~\cite{bui2023rosteals} encodes messages in the latent space using a frozen VQVAE~\cite{esser2021taming}, but imperceptibility is constrained by VQVAE reconstruction quality.

Despite these advancements, current watermarking algorithms face limitations in their effectiveness as a soft-binding solution for AI-generated images. Previous algorithms, often trained and evaluated on low-resolution images, encounter difficulties maintaining their performance when scaled to the higher resolutions, which is now standard in modern image generation models like Stable Diffusion and DALL$\cdot$E 3~\cite{rombach2022high}. Furthermore, the inherent tradeoff between capacity, imperceptibility, and robustness restricts watermark payload capacity, typically to under 100 bits. This limited capacity elevates the risk of ID collisions in the presence of bit errors, compromising the reliability of watermark extraction. Additionally, post-generation watermarking techniques can introduce perceptible artifacts, negatively impacting image quality and hindering adoption, especially in creative fields where visual fidelity is paramount.

To address these challenges, we introduce InvisMark, a novel approach rooted in the insight that high-resolution images inherently possess the capacity to embed a multitude of imperceptible signals. By leveraging carefully crafted neural network architectures and training strategies, we can effectively harness this potential. Our contributions can be summarized as follows:

\begin{table*}
  \centering
  {\small{
  \begin{tabular}{@{}c|c|c|c|c|c|c@{}}
    \toprule
      \makecell{Jpeg Compress. \\
      (min q) }& \makecell{Brightness \\ (bri.) } & \makecell{Contrast \\ (con.)} & \makecell{Saturation \\ (sat.)} & 
      \makecell{Gaussian blur \\ ($k, \sigma$)} & \makecell{Gaussian Noise \\ (std.)} & \makecell{Posterize \\ (bits)}
      \\
    \hline
    50 & 075/1.25 & 0.75/1.25 & 0.75/1.25 & 5, 0.1-1.5 & 0.04 & 4 \\
    \hline
    \hline
    \makecell{ColorJiggle \\ (bri., con., sat., hue)} & \makecell{RGB shift \\ (shift limit)} & \makecell{Flip \\ (prob.)} & \makecell{Rotation \\ (deg.)} & \makecell{RandomErasing \\ (scale, ratio)} & \makecell{Perspective \\ (scale)} & \makecell{RandomResizedCrop \\ (scale, ratio)} \\
    \hline
    0.1, 0.1, 0.1, 0.02 & 0.05 & 1.0 & 0-10.0 & 0.02-0.1, 0.5-1.5& 0.1 & 0.75-1.0, 3/4 - 4/3\\
    \hline
    \bottomrule
  \end{tabular}
  }}
  \caption{Noise settings used in watermark training and evaluation. For JPEG compression, $q$ is the compression factor. For Gaussian blur, $k$ is the kernel size and $\sigma$ is the range for the standard deviation of the kernel. Additionally, for Brightness, Contrast and Saturation, only the lower and upper bound values are used to simulate noise, rather than random values within the range. All noise implementations utilize the Kornia library, except for JPEG compression, which employs the torchvision library.}
  \label{tab:noise_setting}
\end{table*}

\begin{enumerate}
\item \textbf{Novel architecture}: We apply resolution scaling during training and employ robust optimization techniques to enhance decoder resilience against common image transformations with minimal impact on encoded image quality.
\item \textbf{State-of-the-art performance}: InvisMark outperforms existing methods in both imperceptibility and robustness across AI-generated and non-AI-generated image datasets. 

\item \textbf{Larger Payload}: We demonstrate the ability to embed 256 bits of watermarks while maintaining exceptional imperceptibility and robustness, expanding the practical applications of our method in real-world scenarios. 

\end{enumerate}

\section{Methodology}\label{sec:meth}

The goal of image watermarking is to embed a secret message $\omega \in \{0, 1\}^l$ (a binary string of length $l$) within a cover image $x$. Similar to previous approaches, InvisMark utilizes an encoder module $\mathcal{E}$ to embed the message and create the watermarked image $\tilde{x} = \mathcal{E}(x, \omega)$ and a decoder module $\mathcal{D}$ to extract it $\tilde{\omega} = \mathcal{D}(\tilde{x})$.  The primary goal is to minimize the perceptual distortions introduced to the cover image (quantified by image quality loss $\mathcal{L}_q(x, \tilde{x})$) while ensuring the watermark remains detectable even after the image undergoes common transformations (measured by watermark recovery loss $\mathcal{L}_r (\omega, \tilde{\omega})$). 

\subsection{Watermark Encoder}\label{sec:sub_enc}
The watermark, initially a bit array, is transformed via a linear layer into a 3D tensor, We then upsample the tensor and add zero-padding to match the downscaled image dimension. 
This preprocessing step is designed to enhance the watermark's resilience to geometric attacks by controlling its spatial distribution. Additionally, it facilitates the co-embedding of multiple watermarks within a single image, allowing for tailored traceability and security features.

We utilize a MUNIT-based encoder architecture with skip connections to preserve fine image details during watermark encoding~\cite{huang2018multimodal}. Inspired by TrustMark, we also incorporate multiple 1$\times$1 convolution layers as post-processing to maintain high fidelity in the encoded image~\cite{bui2023trustmark}. The encoder outputs the watermarked image's residuals at the low resolution, which are then upscaled to match the original image resolution using a TrustMark-like resolution scaling approach~\cite{bui2023trustmark}, ultimately yielding the final watermarked image.

\subsection{Watermark Decoder}\label{sec:sub_dec}
The decoder $\mathcal{D}$ is a network trained to recover the hidden message from the encoded image $\tilde{w} = \mathcal{D}(\tilde{x}) \in \{0, 1\}^l $. We use ConvNeXT-base  with pre-trained weights as the default decoder. ConvNeXT is a ``modernized" version of the standard ResNet~\cite{he2016deep} toward the design of a vision Transformer and has shown to reach state-of-the-art performance on many computer vision tasks\cite{liu2022convnet}. We replace the last classifier layer with a $l$-dimension sigmoid activated fully connected layer to predict the $l$-bit secret message.  We also explored using ResNet-based networks as decoder. However, we observed lower validation robustness due to unstable training arising from Batch Normalization layers, which is deprecated in the ConvNeXT architecture.

\begin{figure*}
  \centering
   \includegraphics[width=1.0\linewidth]{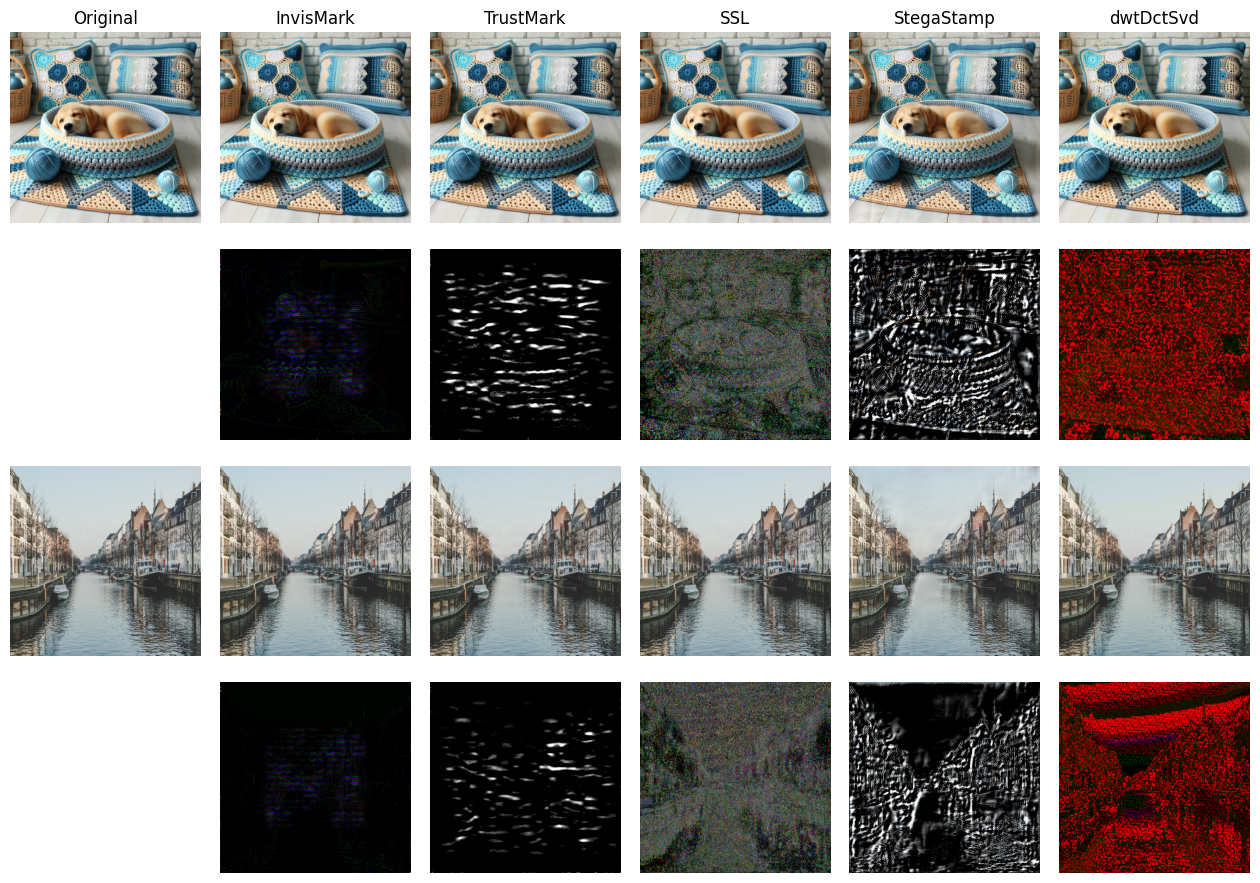}
   \caption{Encoded images and amplified residuals (20$\times$) from InvisMark and 4 other watermarking methods (TrustMark~\cite{bui2023trustmark}, SSL~\cite{fernandez2022watermarking}, StegaStamp~\cite{tancik2020stegastamp} and dwtDctSvd~\cite{navas2008dwt}) on DALL$\cdot$E 3 (top) and DIV2K (bottom) datasets. The residuals from InvisMark are smaller and more imperceptible compared to previous works.}
   \label{fig:wm_example}
\end{figure*}

\subsection{Noiser}\label{sec:sub_noise}
We introduce a noise module after the encoded image. Instead of using randomly sampled noises to enhance robustness, we approach watermark resilience as a robust optimization problem, focusing on worst-case scenarios. Table~\ref{tab:noise_setting} details our noise settings, which are generally aligned with ``medium" noise level settings in TrustMark. However, unlike TrustMark, we introduce variability in geometric transformations to better reflect real-world scenarios. Specifically, for RandomResizedCrop, we randomly crop up to 25\% of the original image area with an aspect ratio ranging from 3/4 to 4/3, then resize it back to the original resolution. For rotation, we apply random rotations of up to 10 degrees. 


\subsection{Model Training}\label{sec:sub_train}
\subsubsection{Dataset}

Our model is trained on 100k DALL$\cdot$E 3 images generated from randomly sampled user prompts. While this training dataset is not available to the public, we evaluate our results against open-source DALL$\cdot$E 3  dataset~\cite{opendalle3}. Additionally, we perform the evaluation on DIV2K, a well-established super-resolution benchmark containing 900 high-quality, 2K resolution images~\cite{Agustsson_2017_CVPR_Workshops}. By benchmarking on both DALL$\cdot$E 3 and DIV2K, we aim to target a more realistic product environment, demonstrating the model's ability to watermark images for both AI-generated and non AI-generated high-resolution images.

\subsubsection{Loss}\label{sec:subsub_train}

The overall training loss is the weighted sum of encoded image quality loss $\mathcal{L}_q$ and watermark recovery loss $\mathcal{L}_r$

\begin{equation}
\mathcal{L} = \alpha_q \mathcal{L}_q (x, \tilde{x})+ \alpha_r \mathcal{L}_r (\omega, \tilde{\omega})
  \label{eq:total_loss}
\end{equation}
where $\alpha_{q, r}$ are the relative weights for each loss term. 

\textbf{Image Quality Loss}: The image quality loss $\mathcal{L}_q$ comprises four components: $\mathcal{L}_{\text{YUV}}$ measures the pixel-level mean squared error in the YUV color space,  $\mathcal{L}_{\text{LPIPS}}$ evaluates the perceptual similarity between the reconstructed and original images~\cite{zhang2018perceptual}, $\mathcal{L}_{\text{FFL}}$ focus on frequency loss targets to reduce artifacts in the frequency domain~\cite{jiang2021focal} and $\mathcal{L}_{\text{GAN}}$ is Wasserstein GAN loss, used to train a discriminator by encouraging the encoded images to be indistinguishable from real images~\cite{arjovsky2017wasserstein}. The $\beta$-coefficients, representing the relative weights of each loss term, are maintained constant 1.0 during training.  Image quality loss is computed at the downscaled resolution rather than the original image resolution.

\begin{equation}
\begin{split}
\mathcal{L}_q (x, \tilde{x}) =
    & \beta_{\text{YUV}}\mathcal{L}_{\text{YUV}} + \beta_{\text{LPIPS}}\mathcal{L}_{\text{LPIPS}} \\
    & + \beta_{\text{FFL}}\mathcal{L}_{\text{FFL}} + \beta_{\text{GAN}}\mathcal{L}_{\text{GAN}}
\end{split}
\end{equation}

\begin{table*}

  \centering
  {\small{
  \begin{tabular}{@{}c|ccccc|ccccc@{}}
    \toprule
      &\multicolumn{5}{c}{DIV2K}  & \multicolumn{5}{c}{DALL$\cdot$E 3}\\
      \hline
      & InvisMark & TrustMark & SSL & StegaStamp & dwtDctSvd & InvisMark & TrustMark & SSL & StegaStamp & dwtDctSvd \\
    \hline
    PSNR & \textbf{51.4} & 41.9 & 42.6 & 37.4 & 38.3 & \textbf{51.4} &  41.9& 42.7 &37.5&40.2\\
    \hline
    SSIM & \textbf{0.998} & 0.995 & 0.989 & 0.978& 0.974 &  \textbf{0.998} & 0.994 & 0.990 &0.976&0.981\\

    \bottomrule
  \end{tabular}
  }}
  \caption{Watermark imperceptibility (quantified by PSNR \& SSIM) for InvisMark and 4 baseline methods. InvisMark consistently achieves highest score for both metrics in DIV2K and DALL$\cdot$E 3 dataset. }
  \label{tab:wm_quality}
\end{table*}

\begin{table*}

  \centering
  {\footnotesize{
  \begin{tabular}{@{}c|ccccc|ccccc@{}}
    \toprule
      &\multicolumn{5}{c}{DIV2K}  & \multicolumn{5}{c}{DALL$\cdot$E 3}\\
      \hline
      & InvisMark & TrustMark & SSL & StegaStamp & dwtDctSvd & InvisMark & TrustMark & SSL & StegaStamp & dwtDctSvd \\
    \hline
    Clean Image&  \textbf{100.0} &  99.7 & 98.6& 99.8 & \textbf{100.0} &  \textbf{100.0} &  99.9 & 97.7 &99.8&99.3\\
    \hline
    Jpeg Compression &  99.5 &  89.7  & 53.9& \textbf{99.8}& 80.8 & 97.5 & 92.9 & 52.6 &\textbf{99.8}&79.6\\
    \hline
    Brightness & \textbf{99.7} & 82.6 & 76.1& 98.0& 93.6& \textbf{99.8} &  84.8 & 74.4 &97.5&92.5\\   
    \hline
    Contrast & \textbf{99.9}  & 81.8 & 71.7& 99.4& 62.8&  \textbf{99.9} & 82.8 & 69.1 &99.1&56.3\\
    \hline
    Saturation & \textbf{100.0} & 94.9 & 89.4& 99.8&59.2&  \textbf{100.0} & 95.7 & 86.6 &99.7&58.3\\
    \hline
     GaussianBlur & \textbf{100.0} & 99.7 & 96.9& 99.8&99.7& \textbf{100.0} & 99.9& 95.5 &99.8&98.2\\
    \hline
     GaussianNoise & \textbf{100.0} & 69.9 & 68.9&88.9 &98.8 & \textbf{100.0} & 65.1 & 68.1 &84.8&96.4\\
    \hline 
     ColorJiggle & \textbf{100.0}  & 89.2 & 69.2& 99.6& 81.4& \textbf{100.0} & 90.5 & 66.7 &99.4&76.3\\
    \hline
     Posterize & \textbf{100.0} & 94.9 & 81.8& 99.7& 99.5& \textbf{100.0} & 96.0 & 80.3 & 99.7 & 98.6\\
    \hline
     RGBShift & \textbf{100.0} & 90.2 & 70.4& 99.8& 53.5&  \textbf{100.0} &  92.5& 67.8 &99.8&47.8\\
     \hline \hline
     Flip & \textbf{100.0}  & 99.7 & 94.8& 50.9& 52.6& \textbf{100.0} & 99.8& 91.4 & 50.7 & 54.3\\
     \hline
     Rotation & \textbf{97.4} & 68.7 & 95.3& 88.2& 52.6& \textbf{98.7} & 73.5& 92.9 &90.8&54.0\\
    \hline
     RandomErasing & \textbf{99.8} & 98.5 & 98.6& 99.5& \textbf{99.8} & \textbf{99.9} &  97.4& 97.6 &99.5&99.1\\
    \hline
      Perspective & \textbf{100.0} & 88.9 & 94.9& 94.4& 50.1& \textbf{100.0} & 90.9 & 92.1 &94.4&50.3\\
    \hline
      RandomResizedCrop & \textbf{97.3} & 96.8 & 92.7& 79.6& 52.4& \textbf{99.8} & 99.7 & 90.6 &81.9&51.7\\
    \bottomrule
  \end{tabular}
  }}
  \caption{Watermark robustness against common image transformations.  InvisMark outperforms all 4 baseline methods for watermark recovery under various image manipulations. Table~\ref{tab:noise_setting} shows the detail of the noise settings used in the evaluation.}
  \label{tab:wm_robust_benchmark}
\end{table*}

\textbf{Watermark Recovery Loss}: In addition to the standard binary cross-entropy loss applied to the encoded image, we also incorporate a watermark recovery loss specifically tailored to images that have undergone a pre-defined list of $n$ image transformations (defined as $\Phi(\tilde{x})$). We treat watermark robustness under various image distortions as a robust optimization problem, focusing on performance across worst-case scenarios. However, to reduce computational complexity, we avoid solving the minimax problem at every iteration. Instead, each 200 steps, we reassess the watermark recovery loss across all noises and select the top-$k$ (with $k=2$) noises that result in the highest loss, as detailed in Equation~\ref{eq:loss_rec}. The noises employed in training remain constant until the next reevaluation.  The parameter $\gamma_i$, controlling the relative weight of the watermark recovery loss after transformation $i$ compared to the original, is held constant at 0.5.

\begin{equation}
\begin{split}
    \mathcal{L}_r (\omega, \tilde{\omega}) =& \mathcal{L}_{\text{BCE}}(\omega, \mathcal{D}(\tilde{x}))  \\
    & + \max_{I \subseteq [0, n], |I| = k}\sum_{i \in I} \gamma_i \mathcal{L}_{\text{BCE}}(\omega, \mathcal{D}(\Phi_i(\tilde{x}))) \\
\end{split}
\label{eq:loss_rec}
\end{equation}

\subsubsection{Training Strategy}\label{sec:subsub_train_strategy}

The training process is structured into three distinct stages:

\textbf{Watermark Extraction}: In the initial stage, we focus on optimizing the watermark decoder by setting the weight for image quality loss $\alpha_q$ to a low value. This approach stabilizes training, as the watermark signal is inherently weaker compared to the image content itself~\cite{bui2023trustmark}.

\textbf{Image Reconstruction}: Once a high watermark recovery rate is achieved, we progressively increase $\alpha_q$ to a predetermined maximum value ($\alpha_{q,max} = 10.0$ in our experiments). This encourages the model to enhance the quality of image reconstruction while maintaining watermark detectability. During the first two stages, we avoid introducing image distortions to enable the model to learn optimal encoding and decoding under ideal conditions. However, due to the end-to-end training with resolution scaling, the model inherently demonstrates robustness against certain distortions like resizing and compression.

\textbf{Robustness Enhancement}: Robust optimization is activated once $\alpha_q$, the weight for image reconstruction loss, reaches its maximum value. We've observed that certain image transformations, such as random cropping and rotation, pose significantly greater challenges for watermark recovery compared to others. Introducing these complex distortions prematurely during training can hinder model convergence. Therefore, we strategically incorporate robust optimization, represented by the second loss term in Equation \ref{eq:loss_rec}, only in the final training stage. This targeted approach allows us to fine-tune the model specifically for resilience against these challenging distortions without compromising performance for other distortions or sacrificing the quality of the encoded image.

\section{Experiments}\label{sec:exp}

We benchmark our approach against recent watermarking and steganography methods, including TrustMark~\cite{bui2023trustmark}, SSL~\cite{fernandez2022watermarking}, StegaStamp~\cite{tancik2020stegastamp}, and dwtDctSvd~\cite{navas2008dwt}. For fair comparison, we use all 900 images from DIV2K dataset and 900 randomly sampled images from open sourced DALL$\cdot$E 3 dataset~\cite{opendalle3}. Each image is encoded with randomly generated 100 bit watermark. We utilize TrustMark-Q and SSL models with default pretrained weights. As no pretrained weights are available for StegaStamp, we retrain the model on our dataset using the default hyperparameters. To evaluate image quality, we employ the standard peak signal-to-noise ratio (PSNR) and Structural Similarity Index Measure (SSIM) for imperceptibility, and decoded bit accuracy for watermark recovery (where 50\% signifies random guessing). All metrics are computed at the original image resolution. We apply resolution scaling if an algorithm lacks native resolution support.

\subsection{Watermark Quality}

\begin{figure}[t]
  \centering
   \includegraphics[width=1.0\linewidth]{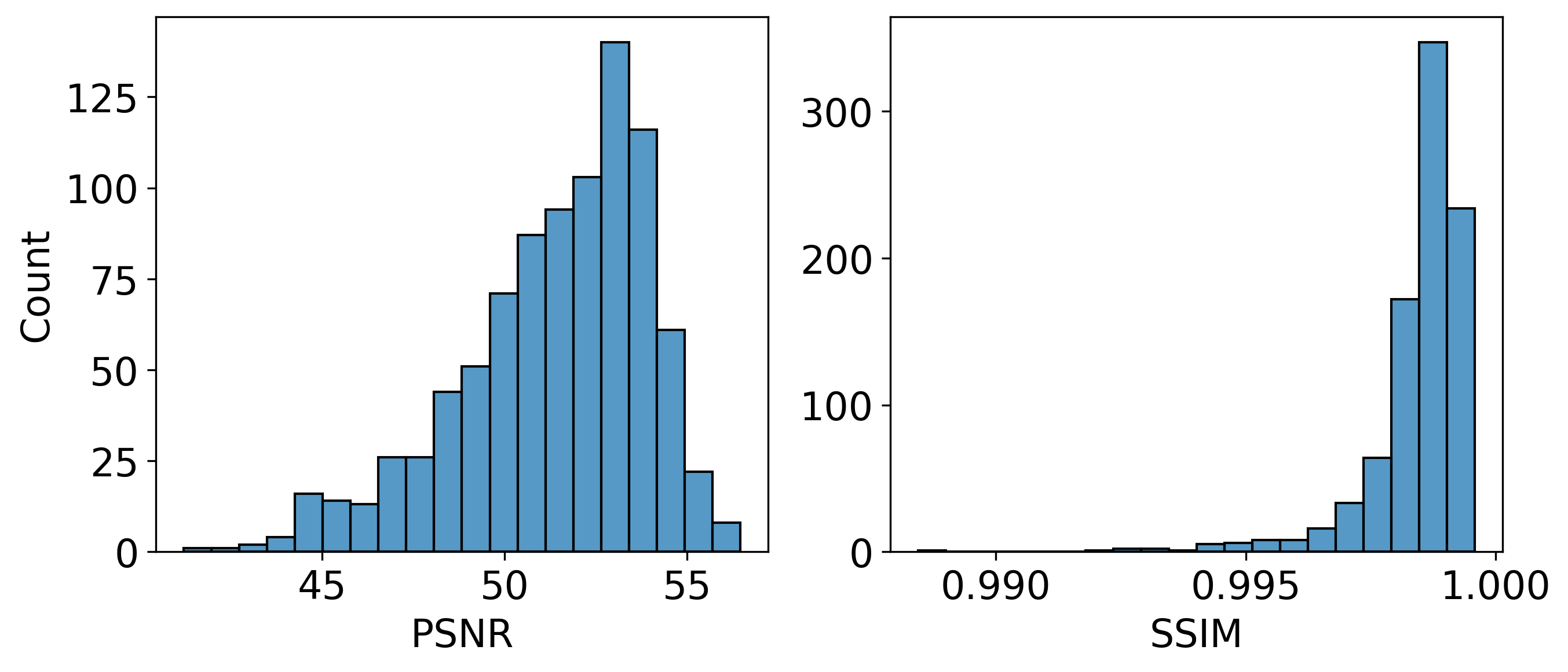}
   \includegraphics[width=1.0\linewidth]{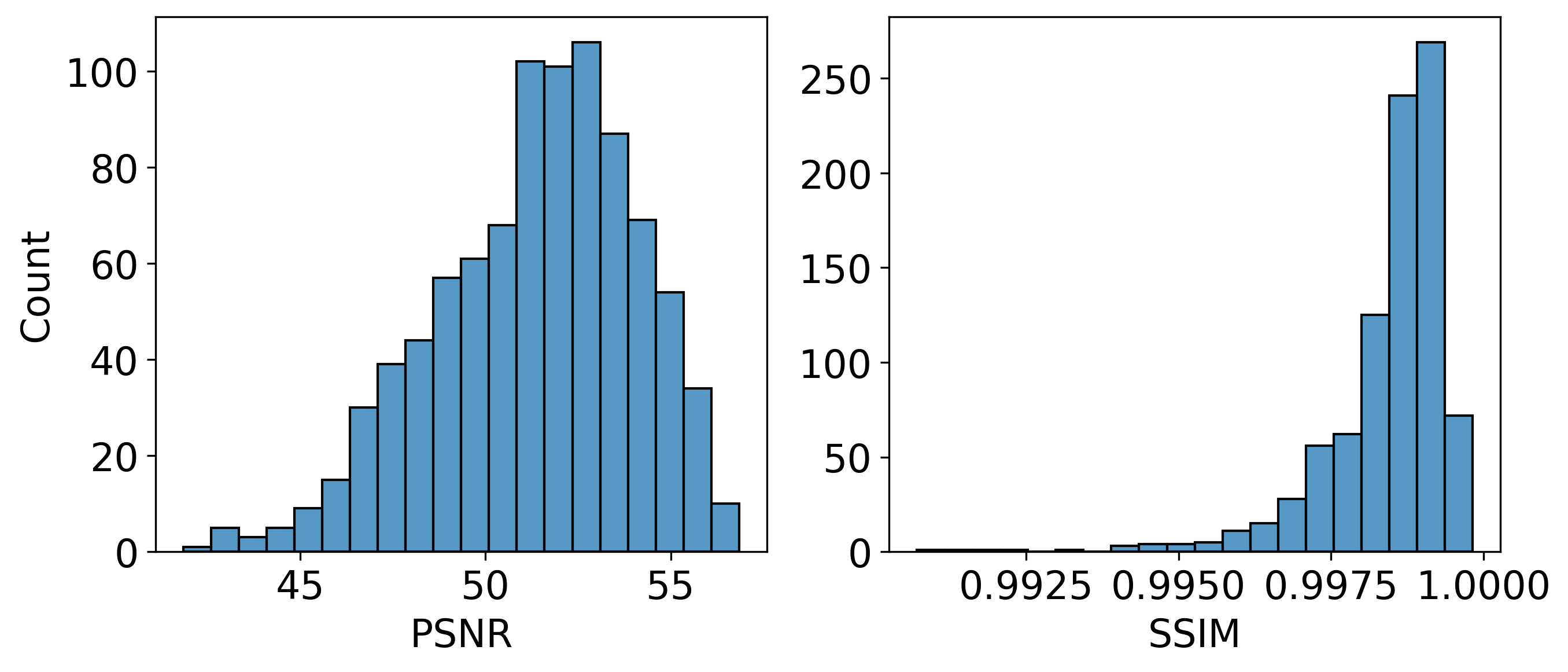}
   \caption{PSNR / SSIM distributions for encoded images from DALL$\cdot$E 3 (top panels) and DIV2K (bottom panels) datasets.}
   \label{fig:exp1}
\end{figure}

Figure \ref{fig:wm_example} presents watermarked images along with their magnified residuals (20$\times$) for both DALL$\cdot$E 3 and DIV2K datasets. TrustMark and StegaStamp generate residuals with a more uniform color distribution, whereas InvisMark exhibits color stripes that could become more noticeable if residual amplitudes were comparable. However, InvisMark's residuals are significantly weaker than other methods, resulting in better image quality and imperceptibility. The watermarked images achieve PSNR values around 51 (refer to Table \ref{tab:wm_quality} and Figure \ref{fig:exp1}), significantly outperforming prior works that reached approximately 40.


\subsection{Watermark Robustness}

A key contribution of this work is the development of several components specifically designed to enhance watermark robustness against common image manipulations. Firstly, by incorporating resolution scaling for watermark residuals directly into the training process, the watermark model becomes inherently resilient to a wide array of transformations, even without additional noise injection. Secondly, we employ a robust optimization technique that prioritizes the most challenging scenarios, ensuring optimal performance against all image transformations encountered during training. Finally, we depart from the commonly used ResNet architecture and employ a more sophisticated decoder model, enabling it to better capture the subtle patterns within encoded watermarks, thereby further boosting their robustness.

Table \ref{tab:wm_robust_benchmark} shows the superior robustness of our watermarking method compared to previous works. To emphasize performance differences, we employed relatively high noise levels, similar to the ``medium" level noises in TrustMark. While all methods perform well on clean images, their effectiveness varies significantly under noisy conditions.

Traditional watermarking techniques like dwtDctSvd typically struggle in the presence of noise, with a few exceptions such as RandomErasing and GaussianBlur. While StegaStamp delivers promising results for high-resolution images when resolution scaling is employed, the encoded image quality often falls short (PSNR$\sim$37),  limiting its creative applications.  Both SSL and TrustMark show reasonable watermark recoverability under noise, but each has specific vulnerabilities: SSL struggles with JPEG compression and ColorShift, while TrustMark is susceptible to Gaussian Noise and Rotation. We were unable to replicate TrustMark's reported results for SSL and StegaStamp, even on clean images. Our analysis indicates that these two methods actually achieve better performance than reported in the TrustMark paper. 

In contrast, InvisMark consistently outperforms the closest baseline across almost all tested image distortions, achieving near-perfect decode bit accuracy for many common noises like Gaussian blur. Even when faced with more challenging distortions, InvisMark maintains remarkable bit accuracy, consistently exceeding 97\%.

\subsection{Watermark Attacks}

Recently, more advanced and sophisticated watermark attacks have emerged. Adversarial attacks, for instance, generate adversarial examples with similar latent space features to deceive watermark detection systems~\cite{saberi2023robustness, lukas2023leveraging}. Regeneration attacks, on the other hand, use diffusion models or VAEs to remove watermarks by introducing noise and subsequently denoising the image~\cite{zhao2023invisible}. Even more concerning are forgery attacks, which aim to replicate and apply legitimate watermarks to unauthorized images\cite{zhu2020secure}. In this section, we will discuss the robustness of our watermarking solution against these attacks and explore potential mitigation strategies.

\begin{figure}[t]
  \centering
   \includegraphics[width=1.0\linewidth]{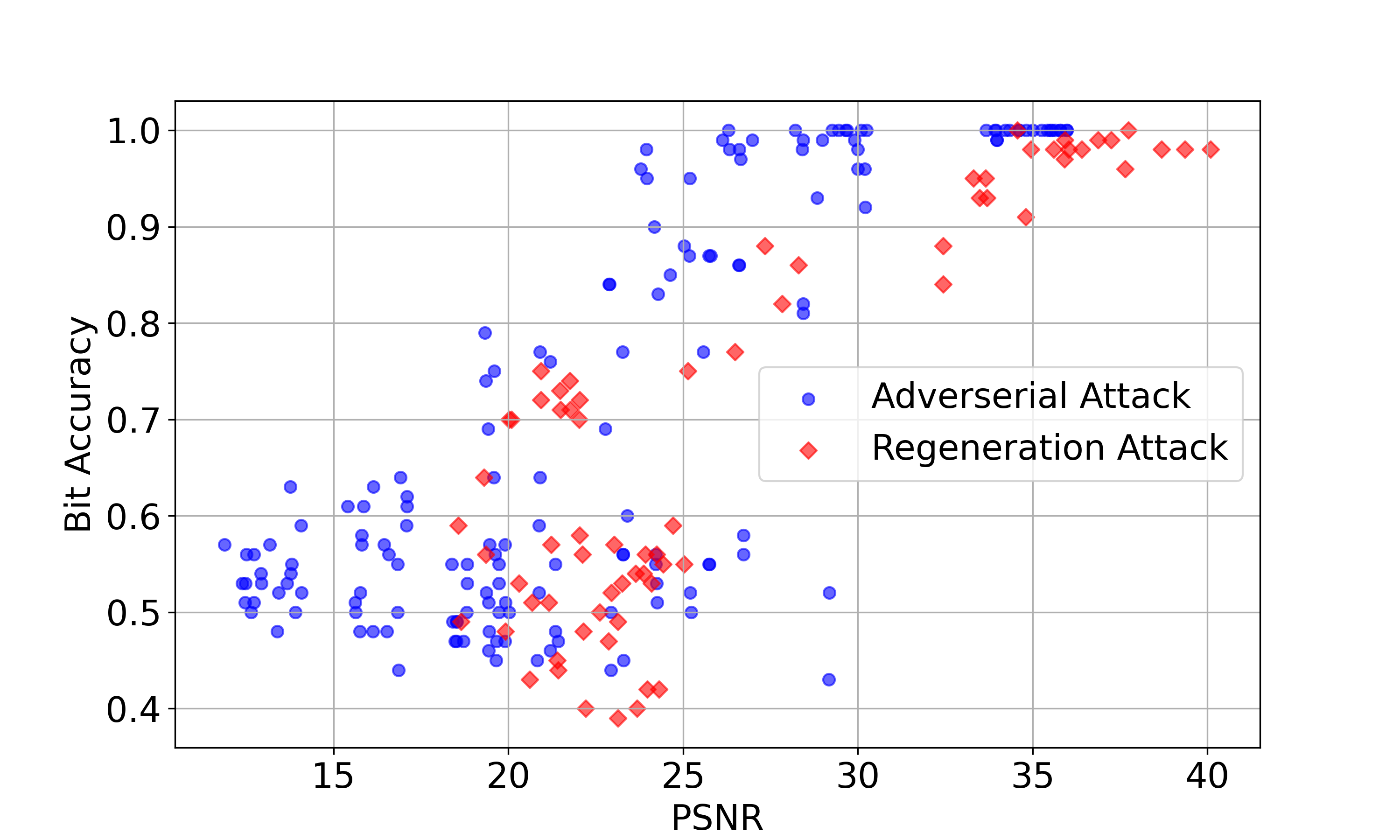}
   \caption{InvisMark's performance under adverserial attack and regeneration attack.}
   \label{fig:attack}
\end{figure}

\textbf{Adverserial Attack:}
We evaluated our watermark solution against embedding attack, a common form of adversarial attack where the goal is to subtly alter the latent representation to remove or disrupt the watermark. To simulate this, we used the off-the-shelf encoder AdvEmbG-KLVAE8, which leverages the KL-VAE encoder (F8)\cite{anwaves}. This grey-box setting mirrors the use of public VAEs in proprietary models, offering a practical test of our watermark's robustness against such sophisticated tampering attempts.

We randomly select 16 different combinations of attack parameters and applied them to 10 sampled images, each embedded with a 100-bit watermark. The resulting bit accuraciy after the attacks are shown in Figure~\ref{fig:attack}. Our watermarking method remains highly robust when the attack does not significantly alter the image content (when PSNR $>$ 30). However, it becomes highly vulnerable between PSNR 25 and 30, and completely removed when PSNR$<$25.

\textbf{Regeneration Attack:}
We also investigated our watermark solution against regeneration attack, specifically via diffusion embedding and reconstruction as discussed in \cite{zhao2023invisible}. Following this work we use the
\textit{stable-diffusion-2-1} from Stable Diffusion\cite{rombach2022high} for the regeneration attack. We change the noising step from [1, 3, 5, 10, 20, 60, 100] on 10 sampled images, each embedded with a 100-bit watermark generated by our approach. Results are shown as red points in  Figure~\ref{fig:attack}. We observe high bit accuracy and high PSNR with relatively low level of attacks. However, the decoding bit accuracy reduces quickly to random guess at PSNR$\sim$25.


\textbf{Forgery Attack and Mitigation}: While regeneration and adversarial attacks can successfully remove our watermarks, they do so at the cost of significantly degrading image quality (PSNR$\sim$25). However, in the context of content provenance, watermark removal merely indicates that watermarking can no longer be used to authenticate images; it does not necessarily imply the image is inauthentic. A more significant threat lies in forgery attacks that could deceive watermark decoders into classifying attacked images as authentic.

We present a simple forgery attack on our watermarks, assuming public access to the watermark encoder. An attacker could extract watermark residuals from pre- and post-watermarked images, then resize and add these residuals to other images to potentially trick the decoder into authenticating the forged content. Table \ref{tab:wm_foregy_attack} demonstrates that both TrustMark and InvisMark decode residuals with high accuracy. However, TrustMark's watermark is lost when residuals are applied to random new images. In contrast, our decoder can still extract watermarks from these new attack images with high accuracy. A subsequent database lookup could then potentially find a valid manifest, incorrectly verifying the image's authenticity.

To mitigate forgery attacks, we propose binding the image watermark to its content using fingerprinting techniques~\cite{collomosse2024authenticity}. Content Credentials enable storing image fingerprints within the C2PA manifest referenced by the watermark identifier. During database lookup, the queried image's fingerprint is compared to the stored one, rejecting any mismatches and thereby enhancing security.

\begin{table}
  \centering
  {\small{
  \begin{tabular}{@{}c|c|c|c@{}}
    \toprule
    & \makecell{Watermarked \\ Image} & \makecell{Residual} & \makecell{Residual + \\ New Image}\\
    \hline
     TrustMark & 99.9 & 93.5 & 67.0 \\
    \hline
    InvisMark & 100.0 & 99.7 & 97.6\\
    \bottomrule
  \end{tabular}
  }}
  \caption{Bit accuracy when decoding the watermarked image, residual only and after adding residual to a random new image.}
  \label{tab:wm_foregy_attack}
\end{table}

\begin{figure*}[t]
  \centering
   \includegraphics[width=1.0\linewidth]{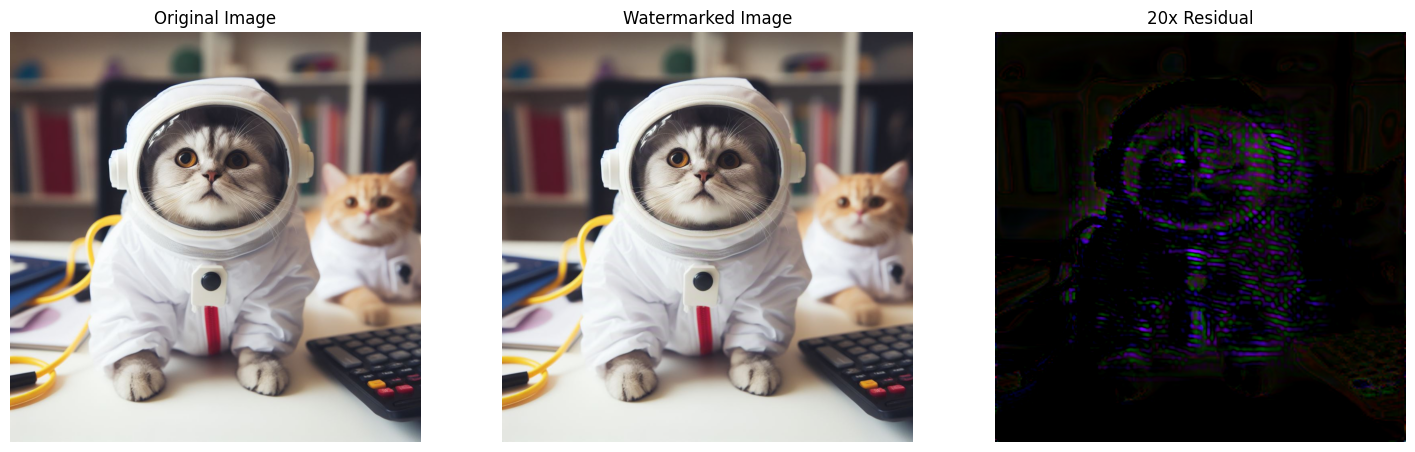}
   \caption{An example image from DALL$\cdot$E 3 dataset with 256-bit encoded watermark.}
   \label{fig:bit256_example}
\end{figure*}

\subsection{UUID encoding}\label{sec:uuid}

While longer watermarks can negatively impact image quality and watermark recovery, our work demonstrates the successful encoding of significantly longer watermarks while still maintaining exceptional image quality and watermark recoverability. Figure \ref{fig:bit256_example} shows an example image with an imperceptible 256-bit watermark. The image quality is further supported by an average PSNR of 48 and SSIM of 0.997 (see Table \ref{tab:wm_uuid_quality}), surpassing prior work with shorter watermarks. InvisMark also exhibit remarkable robustness, achieving an average bit accuracy of 99\% across tested transformations, with a worst-case $\sim$97\% accuracy under resized crop or JPEG compression.

Besides bit accuracy, we introduce watermark decoding success rate as another key metric, representing the probability of complete watermark recovery with error corrections. This is particularly vital for content provenance applications, where accurate decoding of all bits is essential for database retrieval. Employing Bose–Chaudhuri–Hocquenghem (BCH) error correction code~\cite{bose1960class} and 128-bit UUIDs as data bits, we achieve a perfect 100\% success rate across various transformations (refer to Table \ref{tab:wm_uuid}), and maintain over 90\% success rate even under most challenging image distortions.

Larger payloads also enable further system improvements, such as embedding image fingerprints as part of the watermark secret for client-side verification. This approach could potentially eliminate the need for datastore lookups in the case of fingerprint mismatches, thereby boosting overall system efficiency.


\begin{table}

  \centering
  {\small{
  \begin{tabular}{@{}c|c|c@{}}
    \toprule
      & DIV2K & DALL$\cdot$E 3\\
      \hline
    PSNR & 47.9 &  47.8 \\
    \hline
    SSIM & 0.997  & 0.997  \\
    \bottomrule
  \end{tabular}
  }}
  \caption{Image PSNR \& SSIM after encoding 256-bit watermarks.}
  \label{tab:wm_uuid_quality}
\end{table}

\begin{table}
  \centering
  \resizebox{0.99\columnwidth}{!}{
  {\footnotesize{
  \begin{tabular}{@{}c|cc|cc@{}}
    \toprule
      &\multicolumn{2}{c}{DIV2K}  & \multicolumn{2}{c}{DALL$\cdot$E 3}\\
    \hline
      & Bit Acc. & Succ. Rate & Bit Acc. & Succ. Rate \\
    \hline
    \hline 
    Clean Image &  100.0 &  100.0 & 100.0 & 100.0 \\
    \hline
    JPEG Comp. &  99.3 &  98.9 & 97.7 & 92.7 \\
    \hline
    Brightness & 99.3 & 97.6 & 99.6 & 99.0 \\   
    \hline
    Contrast & 99.9 & 99.5 & 99.9 & 99.8 \\
    \hline
    Saturation & 100.0 & 100.0 & 100.0 & 100.0 \\
    \hline
     GaussianBlur & 100.0 & 100.0 & 100.0 & 100.0 \\
    \hline
     GaussianNoise & 100.0 & 100.0 & 100.0 & 99.9 \\
    \hline 
     ColorJiggle & 100.0 & 100.0 & 100.0 & 99.9 \\
    \hline
     Posterize & 100.0 & 100.0 & 100.0 & 100.0 \\
    \hline
     RGBShift & 100.0 & 100.0 & 100.0 & 100.0 \\
     \hline \hline
     Flip & 100.0  & 100.0 & 99.9& 100.0 \\
     \hline
     Rotation & 99.5 & 98.9 & 99.8 & 99.9 \\
    \hline
     RandomErasing & 99.5 & 99.3 & 99.6& 99.8 \\
    \hline
      Perspective & 99.9 & 100.0 & 99.9 & 100.0\\
    \hline
      RandomResizedCrop & 97.9 & 95.2 & 99.7 & 99.4 \\
    \bottomrule
  \end{tabular}
  }}}
  \caption{Watermark robustness assessed under various image transformations after encoding a 256-bit watermark comprising a 128-bit UUID and error correction bits. Success is defined as the accurate decoding of all data bits post-error correction.}
  \label{tab:wm_uuid}
\end{table}

\subsection{Limitations}
\begin{figure}[t]
  \centering
   \includegraphics[width=1.0\linewidth]{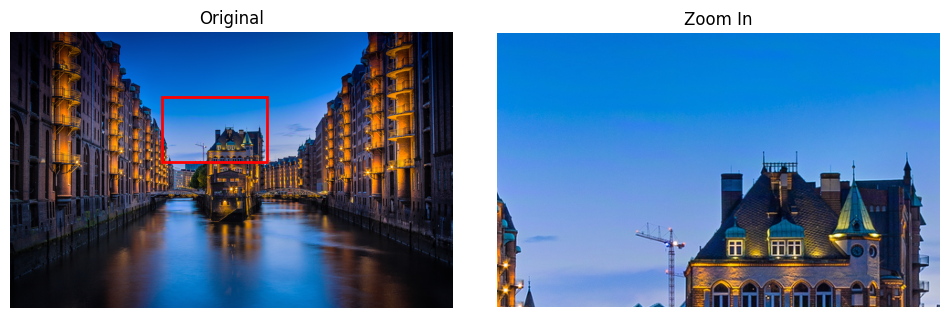}
   \caption{The image on the left shows an example of a watermarked image. On the right, we've zoomed in on the areas marked by red squares in the left panel to highlight subtle artifacts that might be visible due to the watermarking process. }
   \label{fig:wm_perceptiable}
\end{figure}

While InvisMark typically preserves extremely high image quality, perfect imperceptibility cannot be guaranteed. In rare instances, subtle artifacts may become visible in areas with large, uniform backgrounds, like clear blue skies (see Figure~\ref{fig:wm_perceptiable} for examples). Future research focusing on enhancing residual color uniformity could significantly address this issue.

\section{Conclusion}\label{sec:con}

In this work, we presented InvisMark, a novel high-capacity and robust watermarking approach dedicated for high-resolution images. We leverage resolution scaling and robust optimization techniques and significantly improved the decoder's resilience against common image transformations, while minimizing visual artifacts in the encoded image. Through extensive experiments, we demonstrated that InvisMark achieves superior performance in terms of imperceptibility and robustness across diverse datasets, including both AI-generated and traditional images, significantly surpassing existing approaches.  We push the boundaries of watermark payload capacity by successfully embedding 256 bits of watermark while maintaining high levels of imperceptibility and resilience.  We also examined the vulnerabilities of our watermarks, acknowledging that sophisticated attacks can potentially remove our watermarks at the cost of degrading image quality. To address more severe threats such as forgery attacks, we propose integrating InvisMark with complementary fingerprinting techniques. This combined approach effectively mitigates such attack vectors, making InvisMark a trustful tool for ensuring digital content traceability and authenticity.

{\small
\bibliographystyle{ieee_fullname}
\bibliography{egbib}
}

\end{document}